\documentclass[aps,prl,reprint,superscriptaddress]{revtex4-2}

\usepackage{graphicx} 
\usepackage{xcolor} 
\usepackage{amsfonts} 
\usepackage{amssymb} 
\usepackage{amsmath} 
\usepackage[pdftex]{hyperref} 
\hypersetup{colorlinks, 
			linkcolor={blue!75!black!80!yellow},
			citecolor={blue!75!black!80!yellow},
			urlcolor={blue!75!black!80!yellow},
			pdfstartview=FitH}
\usepackage{txfonts}  
\usepackage{txfontsb} 

\usepackage{bm} 
\usepackage[ruled,vlined]{algorithm2e} 

\usepackage{easybmat}

\newcommand{\overbar}[1]{\mkern 2mu\overline{\mkern-2mu#1\mkern-2mu}\mkern 2mu}

\newcommand{\hkuaffil}{\footnotesize Department of Physics and HK Institute of Quantum Science and Technology,\\ The University of Hong Kong, Hong Kong 999077, China}
\newcommand{\pkuaffil}{\footnotesize State Key Laboratory of Advanced Optical Communication Systems and Networks,\\ School of Electronics, Peking University, Beijing 100871, China}
\newcommand{\iopaffil}{\footnotesize Beijing National Laboratory for Condensed Matter Physics and Institute of Physics,\\ Chinese Academy of Sciences, Beijing 100190, China}

\usepackage{textcomp} 
\usepackage{xifthen}
\usepackage{etoolbox}
\newboolean{togglecomments}
\newboolean{togglechanges} 

\setboolean{togglecomments}{false}
\setboolean{togglechanges}{true}

\newcommand{\comment}[2]{%
    \ifbool{togglecomments}%
    {\textcolor{blue!70!black}{\small\textsf{%
    \textsuperscript{\textsc{\textsf{\MakeLowercase{#1}}}}%
    [#2]}}} 
    {}}     
\newcommand{\swap}[2]{\ifbool{togglechanges}
    {#2}  
    {\textcolor{red!70!black}{[#1]}\textrightarrow{}\textcolor{green!50!black}{[#2]}}}
\newcommand{\remove}[1]{\ifbool{togglechanges}
    {}    
    {\textcolor{red!70!black}{#1}}}
\newcommand{\inset}[1]{\ifbool{togglechanges}
    {#1}  
    {\textcolor{green!50!black}{#1}}}
\newcommand{\optional}[1]{\ifbool{togglechanges}
    {#1}  
    {\textcolor{yellow!50!orange!80!gray}{#1}}}
\newcommand{\citeremind}[1]{%
    [\textcolor{blue!75!black!80!yellow}{
        $\blacksquare$%
           \ifthenelse{\isempty{#1}}
               {}
               {\textsuperscript{\textsf{#1}}}%
        }]\xspace}
\newcommand{\todo}[1]{
    \textcolor{orange!80!yellow!95!black}{\textbf{[}%
        \ifthenelse{\isempty{#1}}%
        {\text{$\blacksquare$}}%
        {{\small\textsf{#1}}}%
        \textbf{]}}}

\begin{document}

\title{Efficient algorithms for surface density of states in topological photonic and acoustic systems}

\author{Yi-Xin Sha}
\email{yxsha@hku.hk}
\affiliation{\hkuaffil}
\author{Ming-Yao Xia}
\affiliation{\pkuaffil}
\author{Ling Lu}
\affiliation{\iopaffil}
\author{Yi Yang}
\email{yiyg@hku.hk}
\affiliation{\hkuaffil}

\begin{abstract}
\inset{Topological photonics and acoustics have recently \swap{emerged as promising fields}{garnered wide research interests} for their \swap{robust}{topological} ability to manipulate the light and sound at surfaces. Conventionally, the supercell technique is the standard approach to \remove{studying and} \swap{simulating}{calculating} these boundary effects, whereas it \swap{is constrained by computational resources because of the large system size}{consumes increasingly large computational resources as the supercell size grows}. Additionally, it falls short in differentiating the surface states at opposite boundaries and from bulk states due to the finite size of systems. To overcome the limitations, here we provide two \swap{alternative}{complementary} efficient methods for obtaining the ideal topological surface states of a semi-infinite system. The first one is the cyclic reduction method, which is based on iteratively inverting the Hamiltonian for a single unit cell, and the other is the transfer matrix method, which relies on the eigenanalysis of a transfer matrix for a pair of unit cells. Benchmarks show that, compared to the traditional supercell method, the cyclic reduction method can reduce both memory and time consumption by two orders of magnitude; the transfer matrix method can reduce memory by an order of magnitude, take less than half the time, and \swap{but with higher}{achieve high} accuracy. \swap{They are further extended}{Our methods are applicable} to more complex scenarios, such as coated structures, heterostructures, and sandwiched structures. As examples, the surface-density-of-states spectra of photonic Chern insulators, valley photonic crystals, and acoustic topological insulators are demonstrated. Our computational schemes \inset{enable direct comparisons with near-field scanning measurements and} expedite the exploration of topological artificial materials and the design of topological devices.}
\end{abstract}



\maketitle

\section{\label{sec:I}Main}


Topological photonic and acoustic crystals have emerged as versatile platforms for exploring topological physics, garnering considerable interest these years~\cite{lu2014topological,gangaraj2017berry,ozawa2019topological,davis2021photonic,ma2022topological,zhang2018topological,ma2019topological,xue2022topological,yuan2018synthetic,lustig2021topological,yang2015topological,price2022roadmap,yang2024non}. One of their remarkable features is that their surface states are robust against defects and disorders, which brings potential in realizing useful devices like waveguides, antennas, splitters, and lasers~\cite{chen2019pseudospin,yu2021topological,zhang2021topological,zhou2023protected,zhang2018directional,lumer2020topological,xu2021broadside,wu2023topological,wang2022extended,bandres2018topological,zeng2020electrically}. To study the surface effects, calculating the band structure of a supercell (finite-sized slab) has always been the method of choice, but it has several drawbacks. One problem is that it is hard to distinguish the surface states on both sides of the slab unless the eigenfunctions are calculated and examined~\cite{farmanbar2016green}. Another issue is that the slab thickness should be large enough to avoid the spurious coupling between the surface states at the two boundaries, leading to substantial consumption of computational resources~\cite{smidstrup2017first}. Most importantly, the surface bands are mixed with the bulk bands, and the results cannot be directly compared with the surface state spectrum measured from the near field scanning experiments~\cite{metalidis2005green}.

\inset{The key to solving these problems is the surface Green's function for a semi-infinite system, from which one can derive the surface properties such as surface density state spectrum at a single well-defined boundary. Mathematically, the surface Green's function can be evaluated as the inverse of Hamiltonian with a block-Toeplitz tridiagonal structure~\cite{golub2013matrix}. General direct solvers such as LU and Cholesky factorization scale poorly with increasing system size~\cite{minchev2003some}. Fortunetely, certain algorithms~\cite{velev2004equivalence} can significantly enhance the computational efficiency if the Toeplitz property is fully considered, primarily falling into two categories. One is the iterative technique such as the cyclic reduction method (CRM)~\cite{guinea1983effective,sancho1984quick,sancho1985highly}, and the other is the semi-analytical technique such as the transfer matrix method (TMM)~\cite{lee1981simple1,lee1981simple2,lee1981renormalization,lee1981new,chang1982complex}. Historically, Golub first proposed the CRM for fast calculation of the inverse of a scalar-cyclic operator when solving Poisson equations~\cite{buzbee1970direct}. Then it was extended to deal with the block-cyclic one~\cite{heller1976some} and the semi-infinite one~\cite{sancho1985highly}. Simultaneously, Lee and Joannopoulos put forward the TMM for efficiently inverting the Hamiltonian for Schrödinger equations in semi-infinite systems~\cite{lee1981simple2}, which is similar to the core idea of subsequent developed Mobius transformation method~\cite{umerski1997closed}.}

\inset{These mathematical advances have been successfully transferred to the study in electronic systems. For example, the CRMs in plane-wave and tight-binding bases were used to calculate the electronic transmission of carbon nanotubes and semiconductors~\cite{nardelli1999electronic,smidstrup2017first}. Besides, the TMMs based on tight-biding models were used to investigate the decay of surface states in and graphenes~\cite{rungger2008algorithm,reuter2011probing}, and to image the surface bands of superconductors and topological insulators~\cite{farmanbar2016green,peng2017boundary}.
Currently, these methods have been widely used to analyze and design the surface properties of electronic materials~\cite{cojuhari2009discrete,lu2022defect,schomerus2023renormalization,mustonen2024exact} and even developed as a powerful tool for exploring novel topological phenomena~\cite{wu2018wanniertools}.}

As for photonic and acoustic systems, however, the mathematical advances have not been fully exploited. Although a CRM in a finite element basis is put forward to calculate the surface states of photonic and acoustic topological semimetals~\cite{cheng2020discovering,sha2021surface}, the formulation is limited to the case of a bare semi-infinite structure. On the other hand, a TMM based on a plane-wave basis is presented to simulate the wave propagation in more complex cases such as sandwiched photonic crystals~\cite{li2003light,che2008analysis}, but such non-localized basis functions are hard to describe the optical fields in metallic materials and sound waves in rigid bodies.

In order to address the above limitations, in this work, we implement the CRM and TMM using finite element discretization in photonic and acoustic systems and provide the computational paradigms across a variety of scenarios, such as a bare semi-infinite crystal, a semi-infinite crystal with a surface defect, two semi-infinite crystals interfaced with each other, and two semi-infinite crystals with an interface defect. We demonstrate the utility of our approach by calculating the surface state spectra of photonic Chern insulators, valley photonic crystals, and acoustic topological insulators and compare the differences in the computational efficiency and accuracy between the two methods.

\begin{figure*}[htbp!]
\centering
\includegraphics[width=\linewidth]{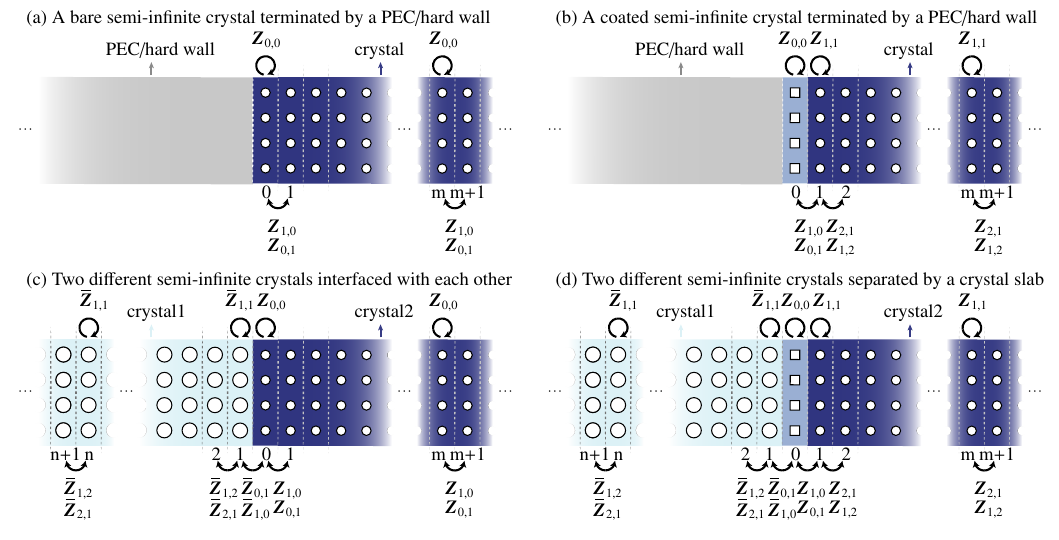}
\caption{\label{fig1}
\textbf{Various photonic or acoustic crystal structures which support topological surface states.} The structure is divided into layers along the direction perpendicular to the surface indicated by an index $m$. ${\bm{Z}_{m,m}}$ is the intra-coupling matrix within a single layer, ${\bm{Z}_{m,m+1}}$ and ${\bm{Z}_{m+1,m}}$ are the inter-coupling matrices between two nearest-neighbor layers. $\overbar{\bm{Z}}$ represents the coupling matrix in the opposite direction. (a) A bare semi-infinite crystal. (b) A semi-infinite coated crystal. (c) Two different semi-infinite crystals interfaced with each other. (d) Two different semi-infinite crystals separated by another crystal slab.}
\end{figure*}

\section{\label{sec:II}Computational framework}

\subsection{\inset{Green's functions and local density of states}}
The Green's functions in photonic [Eq.~\eqref{eq:GF_photonic}] and acoustic [Eq.~\eqref{eq:GF_acoustic}] systems can be defined as the solutions of the wave equations for a point source:
\begin{subequations}
\begin{align}
\begin{split}
\underbrace{ \left[\bm{\nabla}  \times \bm{\mu}^{-1}(\bm{r}) \cdot \bm{\nabla}  \times -{\omega^2} \bm{\varepsilon}(\bm{r}) \right] }_{\bm{Z} \left( \bm{r} ; \omega \right)} \bm{G} \left( \bm{r}, \bm{r}^{\prime}  ; \omega \right)   &= \bm{I}\delta(\bm{r}-\bm{r}^{\prime}),
\label{eq:GF_photonic}
\end{split}\\
\begin{split}
 \underbrace{\left[ \bm{\nabla}  \cdot {\rho}^{-1}(\bm{r}) \bm{\nabla}  + {\omega^2}{K}^{-1}(\bm{r}) \right]}_{\bm{Z} \left( \bm{r} ; \omega \right)}  {G} \left( \bm{r}, \bm{r}^{\prime}  ; \omega \right) &= \delta(\bm{r}-\bm{r}^{\prime}), 
 \label{eq:GF_acoustic}
\end{split}
\end{align}
\label{eq:GF}
\end{subequations}
where $\bm{\mu}(\bm{r})$, $\bm{\varepsilon}(\bm{r})$, $\rho(\bm{r})$ and $K(\bm{r})$ are the permeability, permittivity, mass density and bulk modulus at the location $\bm{r}$, respectively. $\delta(\bm{r}-\bm{r}^{\prime})$ is a Dirac’s delta source at $\bm{r}^{\prime}$. For simplicity, we write the above equations as $\bm{Z}\bm{G}=\bm{I}$, where $\bm{Z}$ combines the differential operators and material parameters, and $\bm{G}$ is the Green's function.

Local density of states (LDOS) describes the spatial distribution of the intensity of a single-particle eigenstate, and can be calculated from the imaginary part of $\bm{G}$~[in Eq.~\eqref{eq:GF}] by imposing an infinitesimal imaginary frequency $\eta$~\cite{economou2006green,novotny2012principles,carminati2015electromagnetic,chew2019green}:
\begin{subequations}
\begin{align}
\begin{split}
{\rm LDOS} \left( {\bm{r}  ;\omega} \right) &=
\frac{2\omega}{\pi }{\rm Im} \left\{ {\rm Tr}\left[ \bm{\varepsilon}\left( \bm{r}\right) \mathop {\lim }\limits_{\eta  \to {0^ + }} \bm{G}\left( \bm{r}, \bm{r}  ; \omega  + \rm{i} \eta \right) \right] \right\},
\label{eq:LDOS photonic}
\end{split}\\
\begin{split}
{\rm LDOS} \left( {\bm{r}  ;\omega} \right) &=
\frac{2\omega}{\pi }{\rm Im} \left[ {K}^{-1}(\bm{r})  \mathop {\lim }\limits_{\eta  \to {0^ - }} G\left( \bm{r}, \bm{r}  ; \omega  + \rm{i} \eta \right) \right].
\label{eq:LDOS acoustic}
\end{split}
\end{align}
\label{eq:LDOS}
\end{subequations}
Here Eq.~\eqref{eq:LDOS photonic} and Eq.~\eqref{eq:LDOS acoustic} are the computation expressions for photonic and acoustic systems, respectively. $\rm Tr$ in Eq.~\eqref{eq:LDOS photonic} signifies tracing the Green's tensor, as it is necessary to consider all polarization degrees of freedom in a photonic system. 

\inset{Because this work mainly focuses on the topological states localized at the system boundaries, we define the surface density of states (SDOS) as the average of LDOS over the surface layer for the following calculations and analyses \inset{unless noted otherwise}}. 

\subsection{\inset{Mathematical origin}}
\inset{Mathematically, our goal is to efficiently find the corner block inverse of the operator $\bm{Z}$ in Eq~\eqref{eq:GF}, and further to obtain the SDOS using Eq.~\eqref{eq:LDOS}.}
\inset{When the system exhibits semi-infinite crystal periodicity [as shown in Fig.~\ref{fig1}(a)] and each crystal layer is discretized identically in finite element method~\cite{jin2015finite}, the operator transforms into the following form:}
\begin{equation}
 \bm{Z}=
\left( \begin{BMAT}{cccc}{ccc}
{{\bm{Z}_{0,0}}}&{{\bm{Z}_{0,1}}}&{}&{}\\
{{\bm{Z}_{1,0}}}&{{\bm{Z}_{0,0}}}&{{\bm{Z}_{0,1}}}&{}\\
{}&{\ddots}&{\ddots}&{\ddots}\\
\end{BMAT} \right),
\label{eq:eigenmatrix0}
\end{equation}
\inset{which is known as block-Toeplitz (block-cyclic) tridiagonal matrix~\cite{golub2013matrix}. Here the diagonal block ${{\bm{Z}_{m,m}}}$ is the intra-coupling within the $m$-th layer, the off-diagonal block ${{\bm{Z}_{m,m+1}}}$/${{\bm{Z}_{m+1,m}}}$ is the inter-coupling between the neighboring layers, and they can both be expressed as $\bm{Z}_{0,0}$ and $\bm{Z}_{0,1}$/$\bm{Z}_{1,0}$, respectively. 
All these blocks are the functions of momenta, considering the periodicity in the closed direction (along the surface).}

\inset{The CRM and TMM are powerful algorithms to invert the structured matrix in Eq.~\eqref{eq:eigenmatrix0}. The CRM is based on iteratively inverting an effective coupling matrix for the surface layer, while the TMM relies on the eigenanalysis of the transfer matrix for a pair of neighboring layers. The methods deal only with the matrices that are the same size or twice the size of $\bm{Z}_{0,0}$, rather than the entire matrix $\bm{Z}$, which significantly reduces the computational resources.}

\inset{This work aims to demonstrate the applicability of these methods to complex photonic and acoustic structures such as those in Fig.~\ref{fig1}(a)-(d) corresponding to Eq.~\eqref{eq:eigenmatrix0}-Eq.~\eqref{eq:eigenmatrix3}, as well as their advantages in conveniently investigating novel topological surface states for direct experimental comparisons. Next, we will introduce CRM and TMM in the context of photonics and acoustics.}



\begin{equation}
\bm{Z} =
\left( \begin{BMAT}{c.cccc}{c.ccc}
{{\bm{Z}_{0,0}}}&{{\bm{Z}_{0,1}}}&{}&{}&{}\\
{{\bm{Z}_{1,0}}}&{{\bm{Z}_{1,1}}}&{{\bm{Z}_{1,2}}}&{}&{}\\
{}&{{\bm{Z}_{2,1}}}&{{\bm{Z}_{1,1}}}&{{\bm{Z}_{1,2}}}&{}\\
{}&{}&{\ddots}&{\ddots}&{\ddots}\\
\end{BMAT} \right)
\label{eq:eigenmatrix1}
\end{equation}

\begin{equation}
\bm{Z}=
\left( \begin{BMAT}{cccc.cccc}{ccc.ccc}
{\ddots}&{\ddots}&{\ddots}&{}&{}&{}&{}&{}\\
{}&{{\overbar{\bm{Z}}_{{1},{2}}}}&{{\overbar{\bm{Z}}_{{1},{1}}}}&{{\overbar{\bm{Z}}_{{2},{1}}}}&{}&{}&{}&{}\\
{}&{}&{{\overbar{\bm{Z}}_{{1},{2}}}}&{{\overbar{\bm{Z}}_{{1},{1}}}}&{{\overbar{\bm{Z}}_{{1},{0}}}}&{}&{}&{}\\
{}&{}&{}&{{\overbar{\bm{Z}}_{{0},{1}}}}&{{\bm{Z}_{0,0}}}&{{\bm{Z}_{0,1}}}&{}&{}\\
{}&{}&{}&{}&{{\bm{Z}_{1,0}}}&{{\bm{Z}_{0,0}}}&{{\bm{Z}_{0,1}}}&{}\\
{}&{}&{}&{}&{}&{\ddots}&{\ddots}&{\ddots}\\
\end{BMAT} \right) 
\label{eq:eigenmatrix2}
\end{equation}


\begin{equation}
\bm{Z}=
\left( \begin{BMAT}{cccc.c.cccc}{ccc.c.ccc}
{\ddots}&{\ddots}&{\ddots}&{}&{}&{}&{}&{}&{}\\
{}&{{\overbar{\bm{Z}}_{{1},{2}}}}&{{\overbar{\bm{Z}}_{{1},{1}}}}&{{\overbar{\bm{Z}}_{{2},{1}}}}&{}&{}&{}&{}&{}\\
{}&{}&{{\overbar{\bm{Z}}_{{1},{2}}}}&{{\overbar{\bm{Z}}_{{1},{1}}}}&{{\overbar{\bm{Z}}_{{1},{0}}}}&{}&{}&{}&{}\\
{}&{}&{}&{{\overbar{\bm{Z}}_{{0},{1}}}}&{{\bm{Z}_{0,0}}}&{{\bm{Z}_{0,1}}}&{}&{}&{}\\
{}&{}&{}&{}&{{\bm{Z}_{1,0}}}&{{\bm{Z}_{1,1}}}&{{\bm{Z}_{1,2}}}&{}&{}\\
{}&{}&{}&{}&{}&{{\bm{Z}_{2,1}}}&{{\bm{Z}_{1,1}}}&{{\bm{Z}_{1,2}}}&{}\\
{}&{}&{}&{}&{}&{}&{\ddots}&{\ddots}&{\ddots}\\
\end{BMAT} \right)
\label{eq:eigenmatrix3}
\end{equation}

\subsection{Cyclic reduction method (CRM)}

\inset{Firstly, let us consider a simple case, a general semi-infinite crystal, as shown in Fig.~\ref{fig1}(a).}
To solve for the surface Green's function $\bm{G}_{0,0}$, we now expand Eq.~\eqref{eq:GF} in a block manner according to Eq.~\eqref{eq:eigenmatrix0}, which gives a series of chain equations:
\begin{subequations}
\begin{align}
\begin{split}
{-\bm{\zeta} _0}{\bm{G}_{m,0}} &= {\bm{\alpha} _0}{\bm{G}_{m + 1,0}} + {\bm{\beta} _0}{\bm{G}_{m - 1,0}}, \; {m} \ge {1},
\label{eq:chain_0p}
\end{split}\\
\begin{split}
{-\bm{\zeta} _0^s}{\bm{G}_{0,0}} &= {\bm{\alpha} _0}{\bm{G}_{1,0}} - \bm{I},
\label{eq:chain_0s}
\end{split}
\end{align}
\label{eq:chain_0}
\end{subequations}
where
\begin{subequations}
\begin{align}
\begin{split}
{\bm{\alpha} _0} &= {\bm{Z}_{0,1}},\; {\bm{\beta} _0} = {\bm{Z}_{1,0}},\; {\bm{\zeta} _0} = {\bm{Z}_{0,0}},
\label{eq:parameter_0p}
\end{split}\\
\begin{split}
\bm{\zeta} _0^s &= {\bm{Z}_{0,0}}.
\label{eq:parameter_0s}
\end{split}
\end{align}
\label{eq:parameter_0}
\end{subequations}
Next, we remove the odd-layer Green's functions repeatedly using Gaussian elimination, and Eq.~\eqref{eq:chain_0}-Eq.~\eqref{eq:parameter_0} are transformed into the following forms after the $i$-th iterations:
\begin{subequations}
\begin{align}
\begin{split}
{-\bm{\zeta} _i}{\bm{G}_{2^i m,0}} &=  {\bm{\alpha} _i}{\bm{G}_{2^i (m + 1),0}} + {\bm{\beta} _i}{\bm{G}_{2^i (m - 1),0}} , \; {m} \ge {1},
\label{eq:chain_mp}
\end{split}\\
\begin{split}
{-\bm{\zeta} _i^s}{\bm{G}_{0,0}} &= {\bm{\alpha} _i}{\bm{G}_{2^i ,0}}- \bm{I},
\label{eq:chain_ms}
\end{split}
\end{align}
\label{eq:chain_m}
\end{subequations}
where
\begin{subequations}
\begin{align}
\begin{split}
&\begin{aligned}
{\bm{\alpha} _{i }} =\ & {\bm{\alpha} _{i-1}}{\left( {{\bm{\zeta} _{i-1}}} \right)^{ - 1}}{\bm{\alpha} _{i-1}}, \\
{\bm{\beta} _{i }} =\ & {\bm{\beta} _{i-1}}{\left( {{\bm{\zeta} _{i-1}}} \right)^{ - 1}}{\bm{\beta} _{i-1}}, \\
{\bm{\zeta} _{i }} =\ & {\bm{\zeta} _{i-1}} - {\bm{\alpha} _{i-1}}{\left( {{\bm{\zeta} _{i-1}}} \right)^{ - 1}}{\bm{\beta} _{i-1}} - \\
&{\bm{\beta} _{i-1}}{\left( {{\bm{\zeta} _{i-1}}} \right)^{ - 1}}{\bm{\alpha} _{i-1}},\\
\end{aligned}
\label{eq:parameter_mp}
\end{split}\\
\begin{split}
&\bm{\zeta} _{i }^s = \bm{\zeta} _{i-1}^s - \bm{\alpha} _{i-1}{\left( {{\bm{\zeta} _{i-1}}} \right)^{ - 1}}\bm{\beta} _{i-1}.
\label{eq:parameter_ms}
\end{split}
\end{align}
\label{eq:parameter_m}
\end{subequations}
Here Eq.~\eqref{eq:chain_m} and Eq.~\eqref{eq:parameter_m} define an effective eigenmatrix which builds connections between the layers at intervals of $2^i$.
As the iteration proceeds, the distance between those layers will increase exponentially, and the corresponding inter-couplings (${\bm{\alpha} _i}$ and ${\bm{\beta} _i}$) will decay exponentially to zero due to the inclusion of a global loss [(i.e. $\eta$ in Eq.~\eqref{eq:LDOS})].

Finally, the surface Green's function $\bm{G}_{0,0}$ decouples with the bulk one $\bm{G}_{2^i,0}$ in Eq.~\eqref{eq:chain_ms}, and we achieve its approximation
\begin{equation}
{\bm{G}_{0,0}} = \mathop {\lim }\limits_{i \to \infty } {\left( {\bm{\zeta}} _i^s \right)^{ - 1}},
\label{eq:CRM SGF}
\end{equation}
which in turn allows for the derivation of the SDOS through Eq.~\eqref{eq:LDOS}. 

Aside from the bare semi-infinite scenario described above, the CRM can also handle other more complicated situations like (line defects between) heterostructures, as shown in Fig.~\ref{fig1}(b)-(d). Their associated pseudo-codes can be found in Algorithm.~\ref{algorithm: CRM coated}-~\ref{algorithm: CRM sandwich}.

\subsection{Transfer matrix method (TMM)}
\inset{Complementary to the CRM, in the following, we describe the TMM for SDOS, which is of higher accuracy at the cost of reduced computational speed. 
We also choose the bare semi-infinite structure as the example for elaboration. Firstly, we relate the Green's functions of each layer by introducing a transfer matrix $\bm{T}$, which also corresponds to Eq.~\eqref{eq:chain_0}:}
\begin{subequations}
\begin{align}
\begin{split}
\begin{pmatrix}
\bm{G}_{m+1,0}\\
\bm{G}_{m,0}\\
\end{pmatrix}
&=
\bm{T}^{m}
\begin{pmatrix}
\bm{G}_{1,0}\\
\bm{G}_{0,0}\\
\end{pmatrix}, \; {m} \ge {1},
\label{eq:TM for Green's function p}
\end{split}\\
\begin{split}
{-\bm{Z}_{0,0}}{\bm{G}_{0,0}} &= {\bm{Z}_{0,1}}{\bm{G}_{1,0}}- \bm{I},
\label{eq:TM for Green's function s}
\end{split}
\end{align}
\label{eq:TM for Green's function}
\end{subequations}
with
\begin{equation}
\bm{T}=
\begin{pmatrix}
-{\bm{Z}_{0,1}^{-1}} {\bm{Z}_{0,0}} & -{\bm{Z}_{0,1}^{-1}} {\bm{Z}_{1,0}}\\
\bm{I}&\bm{0}\\
\end{pmatrix}.
\label{eq:TM}
\end{equation}
Next, we rewrite Eq.~\eqref{eq:TM for Green's function p} in the following form
\begin{equation}
\begin{pmatrix}
\bm{G}_{m+1,0}\\
\bm{G}_{m,0}\\
\end{pmatrix}
=\bm{S} \bm{\Lambda}^{m} \left[ \bm{S}^{-1}
\begin{pmatrix}
\bm{G}_{1,0}\\
\bm{G}_{0,0}\\
\end{pmatrix}
\right],
\label{eq:TM for Green's function through eigenanalysis}
\end{equation}
where $\bm{\Lambda}$ is a diagonal matrix and $\bm{S}$ is a full matrix consisting of all the eigenvalues and eigenvectors of $\bm{T}$, respectively:
\begin{equation}
    \bm{T}\bm{S}=\bm{S}\bm{\Lambda}.
    \label{eq:eigenanalysis}
\end{equation}

Upon analyzing Eq.~\eqref{eq:TM for Green's function through eigenanalysis}, it becomes imperative to eliminate the eigenvalues of $\bm{\Lambda}$ with moduli greater than 1 to avoid divergence in the Green's functions ($\bm{\Lambda}^m=\bm{\infty}$). Consequently, the term enclosed within the square brackets must satisfy the following condition:
\begin{equation}
\bm{S}^{-1}
\begin{pmatrix}
\bm{G}_{1,0}\\
\bm{G}_{0,0}\\
\end{pmatrix}
=
\begin{pmatrix}
\bm{C}\\
\bm{0}\\
\end{pmatrix},
\label{eq:convergence condition}
\end{equation}
where $\bm{C}$ is a constant matrix, $\bm{0}$ is a zero matrix, and their positions correspond to those of eigenvalues with moduli less than 1 and greater than 1 in $\bm{\Lambda}$, respectively. Then $S$ can also be arranged as a partition matrix corresponding to the same eigenvalue distribution in $\bm{\Lambda}$:
\begin{equation}
\bm{S}=
\begin{pmatrix}
\bm{S}_2 & \bm{S}_4\\
\bm{S}_1 & \bm{S}_3\\
\end{pmatrix}.
\label{eq:partition form of S}
\end{equation}
Substituting Eq.~\eqref{eq:partition form of S} into Eq.~\eqref{eq:convergence condition}, we have a relationship between the surface Green's function $\bm{G}_{0,0}$ and the bulk one $\bm{G}_{1,0}$:
\begin{equation}
\begin{aligned}
&\left( {\begin{array}{cc}
\bm{G}_{1,0}\\
\bm{G}_{0,0}\\
\end{array}} \right)
=
\bm{S}
\left( {\begin{array}{cc}
\bm{C}\\
\bm{0}\\
\end{array}} \right)
=
\left( {\begin{array}{cc}
\bm{S}_2\bm{C}\\
\bm{S}_1\bm{C}\\
\end{array}} \right)\\
&\Longrightarrow \bm{G}_{1,0}=\bm{S}_2 \bm{S}_1^{-1} \bm{G}_{0,0}.
\end{aligned}
\label{eq:relation between GF and BG}
\end{equation}

Finally, combining Eq.~\eqref{eq:relation between GF and BG} with Eq.~\eqref{eq:TM for Green's function s}, we obtain an explicit expression for the surface Green's function:
\begin{equation}
\bm{G}_{0,0}=\left( {\bm{Z}_{0,0}} + {\bm{Z}_{0,1}} \bm{S_2} \bm{S_1}^{-1} \right) ^{-1},
\label{eq:TMM SGF}
\end{equation}
and the SDOS for the semi-infinite system can be accordingly derived via Eq.~\eqref{eq:LDOS}. 

One potential difficulty may arise when the inverse of the inter-coupling matrix $\bm{Z}_{0,1}$ in Eq.~\eqref{eq:TM} does not exist. To overcome the problem, one can decompose the transfer matrix in the following way:
\begin{equation}
\bm{T}=\bm{T_1^{-1}}\bm{T_2}=
\left( {\begin{array}{cc}
\bm{0}&\bm{I}\\
-\bm{Z}_{0,1}&\bm{0}\\
\end{array}} \right)^{-1}
\left( {\begin{array}{cc}
\bm{I}&\bm{0}\\
\bm{Z}_{0,0}&\bm{Z}_{1,0}\\
\end{array}} \right),
\end{equation}
and transform the standard eigenvalue problem [see Eq.~\eqref{eq:eigenanalysis}] into a generalized eigenvalue problem to find the eigensolutions of $\bm{T}$~\cite{lee1981renormalization}.

The detailed pseudo-codes of using TMM to handle other complex structures, as shown in Fig.~\ref{fig1}(b)-(d), are summarized in Algorithm.~\ref{algorithm: TMM coated}-~\ref{algorithm: TMM sandwich}.

\section{\label{sec:III}Numerical results}

\begin{figure*}[htbp!]
\centering
\includegraphics[width=0.95\linewidth]{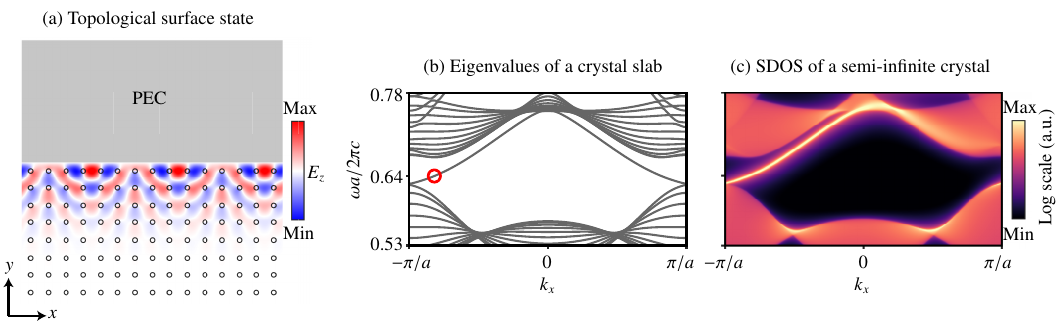}
\caption{\label{fig2}\textbf{Topological surface states of a bare semi-infinite photonic Chern insulator terminated by a PEC.} (a) Mode profile calculated at a normalized frequency of $0.64 c/a$, which corresponds to the red circle in (b). (b) Band structure of a 12-cell photonic Chern insulator slab with two PEC boundaries. (c) SDOS spectrum of a semi-infinite photonic Chern insulator with a PEC boundary. \inset{The SDOS spectrum clearly shows that only a unidirectional chiral edge state can be excited at the PEC-crystal interface, whereas the eigenvalue spectrum presents chiral edge states at opposite boundaries simultaneously.}}
\end{figure*}

\begin{figure*}[htbp!]
\centering
\includegraphics[width=0.95\linewidth]{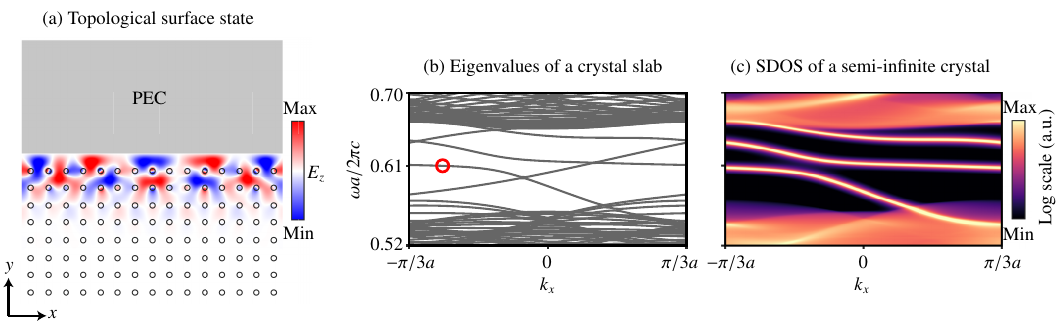}
\caption{\label{fig3}\textbf{Topological surface states of a semi-infinite photonic Chern insulator coated with a crystal slab and terminated by a PEC.} (a) Mode profile calculated at a normalized frequency of $0.61 c/a$, which corresponds to the red circle in (b). (b) Band structure of a 13-cell photonic Chern insulator slab with two PEC boundaries. (c) SDOS spectrum of a semi-infinite photonic Chern insulator with a PEC boundary. \inset{The SDOS spectrum clearly shows that the modified photonic crystal boundary can support a helical topological slow light, while the eigenvalue spectrum redundantly exhibits other surface states on the opposite boundary.}
}
\end{figure*}

\begin{figure*}[htbp!]
\centering
\includegraphics[width=0.95\linewidth]{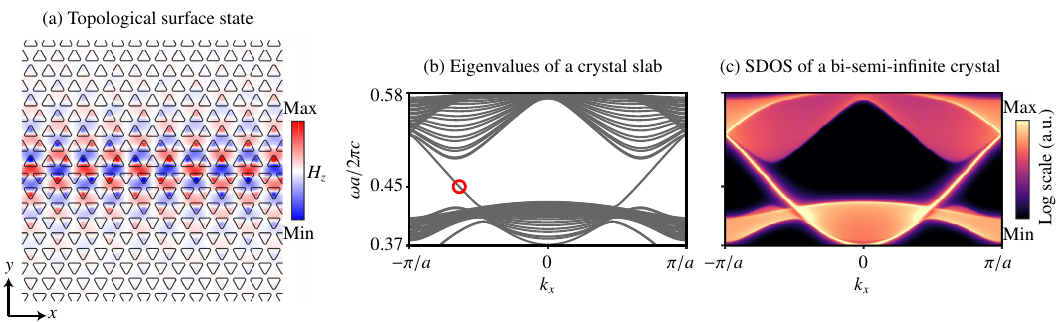}
\caption{\label{fig4}\textbf{Topological surface states of a valley photonic crystal formed by two semi-infinite crystals face to face.} (a) Mode profile calculated at a normalized frequency of $0.45 c/a$, which corresponds to the red circle in (b). (b) Band structure of a 24-cell valley photonic crystal slab with two PEC boundaries. (c) SDOS spectrum of a valley photonic crystal extending infinitely on both sides. \inset{The SDOS spectrum clearly shows that a pair of pseudospin-valley locked topological states are equally exited at the interface of the photonic crystal heterostructure, while the eigenvalue spectrum redundantly presents other trivial states \inset{(at the bottom of b)} on the supercell boundaries.}}
\end{figure*}

\begin{figure*}[htbp!]
\centering
\includegraphics[width=0.95\linewidth]{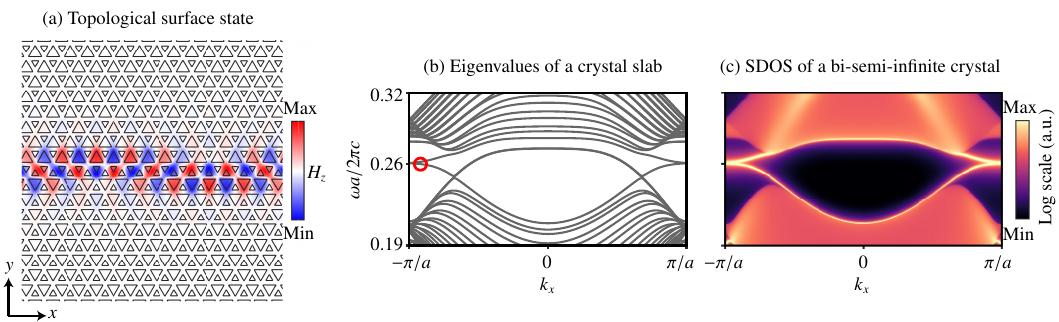}
\caption{\label{fig5}\textbf{Topological surface states of a valley photonic crystal formed by two semi-infinite crystals separated by another crystal slab.} (a) Mode profile calculated at a normalized frequency of $0.26 c/a$, which corresponds to the red circle in (b). (b) Band structure of a 25-cell valley photonic crystal slab with two PEC boundaries. (c) SDOS spectrum of a valley photonic crystal extending infinitely on both sides. \inset{The SDOS spectrum clearly shows that the modified interface can support a pair of topological slow light in different valleys, whereas the eigenvalue spectrum redundantly shows other trivial states on the supercell boundaries.}}
\end{figure*}

\begin{figure*}[htbp!]
\centering
\includegraphics[width=0.95\linewidth]{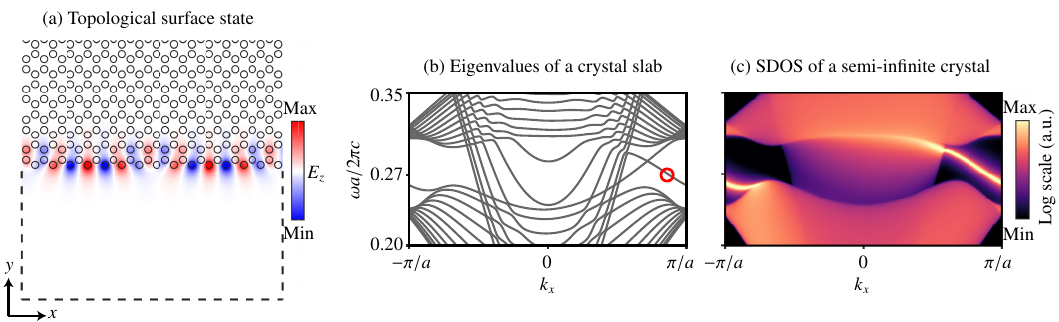}
\caption{\label{fig6}\textbf{Topological surface states of a bare semi-infinite photonic Chern insulator exposed to air.} (a) Mode profile calculated at a normalized frequency of $0.27 c/a$, which corresponds to the red circle in (b). (b) Band structure of a 12-cell photonic Chern insulator slab with an PEC and an open boundary. (c) SDOS spectrum of a photonic Chern insulator with an open boundary. \inset{The SDOS spectrum clearly shows the air-crystal interface can support an unidirectional topological state and its behaviors inside and outside the light cone, whereas the eigenvalue spectrum redundantly shows the states at the PEC-crystal interface.}}
\end{figure*}

\begin{figure*}[htbp!]
\centering
\includegraphics[width=0.95\linewidth]{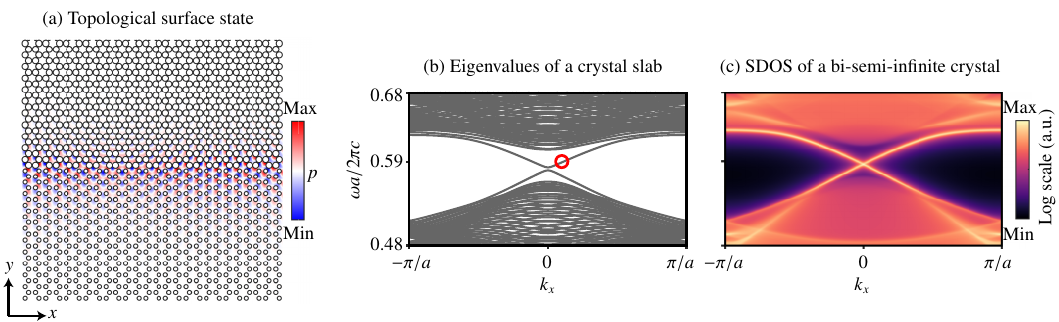}
\caption{\label{fig7}\textbf{Topological surface states of an acoustic topological insulator formed by two semi-infinite crystals face to face.} (a) Mode profile calculated at a normalized frequency of $0.59 c/a$, which corresponds to the red circle in (b). (b) Band structure of a 24-cell acoustic topological insulator slab with two hard wall boundaries. (c) SDOS spectrum of an acoustic topological insulator extending infinitely on both sides. \inset{The SDOS spectrum clearly shows that two pseudospin states are equally excited at the interface of the acoustic crystal heterostructure, while the eigenvalue spectrum makes it difficult to distinguish between surface and bulk states due to the lack of intensity information.} }
\end{figure*}

To showcase the applicability of the CRM and TMM in different complex scenarios, we select four representative photonic structures (see Fig.~\ref{fig2}-Fig.~\ref{fig5}) for verification, which correspond to the four cases illustrated in Fig.~\ref{fig1}(a)-(d). \inset{Moreover, we include another two structures to demonstrate that the developed methodology is equally effective for 1) semi-infinite material/media boundaries (Fig.~\ref{fig6}); and 2) inter-crystal interfaces (Fig.~\ref{fig7}). The latter example also proves its applicability in acoustics.} For all of these six examples, we assume continuous translational invariance along the $z$ direction with $k_z=0$, and introduce the same imaginary frequency $\eta=\omega/1000$ into the original Hermitian systems for a proper broadening of SDOS. In the following, we will give the detailed explanations of these examples.




The first example is a two-dimensional (2D) gyromagnetic photonic Chern insulator with a perfect electric conductor (PEC) cladding [Fig.~\ref{fig1}(a)], which exhibits topological surface states in the second band gap. Fig.~\ref{fig2}(a) presents a schematic of the structure, and the radii of the dielectric pillars are $0.13a$, with the relative permittivity $\bm{\varepsilon}_r$ of $13$ and permeability $\bm{\mu}_r$ of 
$\left( {\begin{array}{ccc}
1     & -0.4\rm{i} & 0   \\
0.4\rm{i} & 1    & 0   \\
0     & 0    & 1   \\
\end{array}} \right)$. We calculate the surface band structure for a 12-cell crystal slab and the SDOS for a bare semi-infinite crystal, which are displayed in Fig.~\ref{fig2}(b) and Fig.~\ref{fig2}(c), respectively. It can be clearly seen that the surface states become more pronounced against the bulk states, and only the chiral state on a single surface is retained in the SDOS spectrum.

Our second example is again, a 2D photonic Chern insulator, but with a surface modification and a PEC cladding [Fig.~\ref{fig1}(b)], which features topological slow light in the second band gap. Fig.~\ref{fig3}(a) is a schematic of the structure, where the radii of the dielectric pillars are $0.15a$, with the relative permeability $\bm{\mu}_r$ of 
$\left( {\begin{array}{ccc}
0.83   & -0.42\rm{i} & 0   \\
0.42\rm{i} & 0.83  & 0   \\
0      & 0     & 1   \\
\end{array}} \right)$. The relative permittivity $\bm{\varepsilon}_r$ of the pillars are $\{15,15,25\}$ for adjacent columns along the $x$ direction while kept uniform along the $y$ direction.
The surface band structure for a 13-cell crystal slab and the SDOS for a semi-infinite coated crystal are calculated and shown in Fig.~\ref{fig3}(b) and Fig.~\ref{fig3}(c), respectively. It can be observed that only topological states with negative group velocities are present in the SDOS spectrum, and they exhibit minimal group velocities at the Brillouin zone boundary.

The third example is a 2D valley photonic-crystal heterostructure [Fig.~\ref{fig1}(c)], which supports valley-dependent surface states in the second band gap~\cite{ma2016all}. Fig.~\ref{fig4}(a) provides a schematic of the structure, which is composed of two types of dielectric pillars with opposite orientations. The pillar has a side length of $0.615a$ and a chamfer length of $0.11a$ at the corner, with the relative permittivity $\bm{\varepsilon}_r$ of $13$ and permeability $\bm{\mu}_r$ of 1. We calculate the surface band structure for a 24-cell crystal slab and the SDOS for a heterostructure infinitely extending into both bulks, which are displayed in Fig.~\ref{fig4}(b) and Fig.~\ref{fig4}(c), respectively. It becomes apparent that a pair of topological states with opposite group velocities are locked in different valleys in the SDOS spectrum.

The fourth example is again, a valley photonic-crystal heterostructure, but with a line defect sandwiched at the interface [Fig.~\ref{fig1}(d)], which supports valley-dependent slow light in the first band gap~\cite{yoshimi2020slow}. Fig.~\ref{fig5}(a) offers a schematic of the structure, which consists of three types of dielectric crystal slabs. The crystals have a background material relative permittivity $\bm{\varepsilon}_r$ of $11.56$ and permeability $\bm{\mu}_r$ of $1$, with the long and short side of air pillars being $1.3a/\sqrt{3}$ and $0.7a/\sqrt{3}$. The surface band structure for a 25-cell crystal slab and the SDOS for an sandwiched structure extending infinitely in both directions are calculated and shown in Fig.~\ref{fig5}(b) and Fig.~\ref{fig5}(c), respectively. It can be intuitively observed that only the confined modes of the beared interface exist in the SDOS spectrum, and it shows slow group velocities near the Brillouin zone edge.

The fifth example is a 2D photonic Chern insulator exposed to air, which has topological surface states in the first band gap~\cite{poo2011experimental}. Fig.~\ref{fig6}(a) is a schematic of the structure, and the radii of the dielectric pillars are $0.2a$, with the relative permittivity $\bm{\varepsilon}_r$ of $15.26$ and permeability $\bm{\mu}_r$ of 
$\left( {\begin{array}{ccc}
0.80   & -0.72\rm{i} & 0   \\
0.72\rm{i} & 0.80  & 0   \\
0      & 0     & 1   \\
\end{array}} \right)$. We calculate the surface band structure for a 12-cell crystal slab and the SDOS for a bare semi-infinite crystal interfaced with air, which are shown in Fig.~\ref{fig6}(b) and Fig.~\ref{fig6}(c), respectively. It is noted that only the state of the air-crystal interface remains in the SDOS spectrum. Moreover, information absent in the surface band calculation can be found in the SDOS spectrum: when the topological surface band enters the light cone, it couples to the radiation continuum (in the $x-y$ plane), resulting in their hybridization and the blurring of the SDOS of the surface band.

The last example is a 2D acoustic topological insulator, which has a pseudospin-dependent topological surface states in the second band gap~\cite{he2016acoustic}. Fig.~\ref{fig7}(a) is a schematic of the structure, and the radii of the rigid bodies on the two sides are $0.3a$ and $0.45a$. We calculate the surface band structure for a 24-cell crystal slab and the SDOS for a heterostructure extending infinitely on both sides, which are shown in Fig.~\ref{fig7}(b) and Fig.~\ref{fig7}(c), respectively. It can be seen from the SDOS spectrum that two pseudospin-locked counterpropagating acoustic modes are equally excited at the same interface.

To summarize these examples, we can see that the efficient calculation of SDOS provides three unique advantages that complement surface band calculations. First, in SDOS spectra, bulk states become genuinely continuous while surface states remain discrete, allowing for a clear visualization of the evolution of surface states, and, in particular, their behaviors within the continuum in frequency-momentum space. Second, the topological states of a single well-defined surface can be obtained directly from SDOS, mitigating the need to inspect individual wavefunctions of surface bands to sort out the localization on different surfaces. Third, the efficient calculation of SDOS spectra has an experimental advantage: they can be directly compared with observables in near-field scanning experiments in both photonics and acoustics.

\begin{figure*}[htbp!]
\centering
\includegraphics[width=0.95\linewidth]{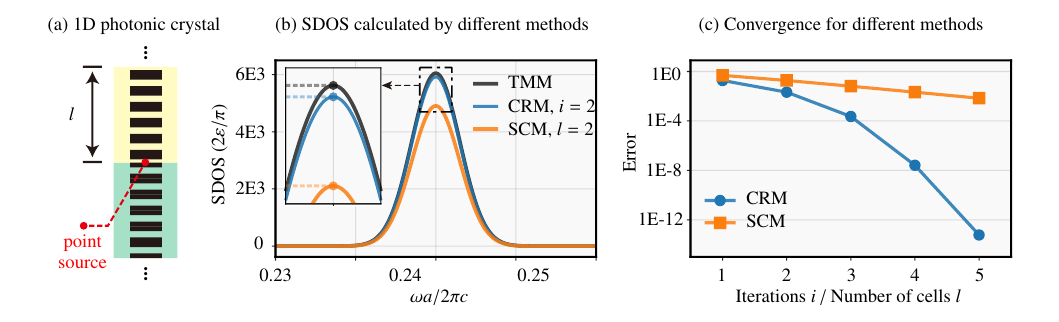}
\caption{\label{fig8}\inset{\textbf{Accuracy and convergence analyses of CRM, TMM, and the conventional supercell method (SCM).} (a) A 1D photonic crystal with inversion symmetry which supports topological interface states. The excitation and receiver are both located at the same point on the interface. (b) The SDOS for a topological state (at the normalized frequency of 0.2425) obtained from different methods. The imaginary frequency is taken as $\eta=\omega/100 $ for finite line broadening. The number of iterations $i$ in the CRM and the number of unit cells $l$ (on one side) in the SCM are both set to 2, and the unknowns $N$ in a unit cell is $\sim$$2000$. It can be observed that the accuracy of TMM is higher than that of CRM, and both are superior to the SCM. (c) The convergence for different methods. The errors are defined as $({\rm SDOS}_i-{\rm SDOS}_{i-1})/{\rm SDOS}_\infty$ for the CRM and $({\rm SDOS}_l-{\rm SDOS}_{l-1})/{\rm SDOS}_\infty$ for the SCM. It can be seen that the error decreases exponentially for the CRM and linearly for the SCM, indicating that the CRM has better convergence.}}
\end{figure*}

\section{\label{sec:IV}{Computing accuracy and efficiency}}
Both the CRM and TMM can effectively obtain the surface Green's function, owing to the block-cyclic tridiagonal form of the matrix $\bm{Z}$ in such periodic systems. In the following, we will evaluate and compare the strengths and weaknesses of the CRM and TMM, particularly in terms of \inset{computing accuracy and efficiency}.

In terms of computational accuracy, the TMM provides higher precision than the CRM. As the TMM directly provides an exact expression for the surface Green's function [see Eq.~\eqref{eq:TMM SGF}], its results are more accurate compared to the iterative approach used by the CRM [see Eq.~\eqref{eq:CRM SGF}]. \inset{To illustrate this, we take a one-dimensional (1D) photonic crystal with inversion symmetry~\cite{vaidya2023topological} as an example to analyze the accuracy and convergence of the CRM, TMM, and the conventional supercell method (SCM) in Fig.~\ref{fig8}.} However, it is important to note that the introduction of imaginary frequency $\eta$ may lead to singularities in the system [such as exceptional points (EPs)]. In this case, $\bm{T}$ becomes ill-conditioned, which could result in catastrophic round-off error amplification, especially at the edges of the energy bands. One possible approach is to reconstruct the relationship between Green's functions and LDOS [see Eq.~\eqref{eq:LDOS}] at EPs~\cite{pick2017general} so that the aforementioned efficient algorithms can be utilized as usual. 

In terms of computational memory, CRM is more advantageous than TMM. The TMM's memory consumption primarily comes from the eigenanalysis of matrix $\bm{T}$, while the CRM's originates from inverting the self-coupling matrix $\bm{\zeta}^s$. Notably, $\bm{T}$ has twice the degrees of freedom as $\bm{\zeta}^s$ [see Eq.~\eqref{eq:parameter_0s} and Eq.~\eqref{eq:TM}], leading to larger memory usage in TMM. As shown in Table~\ref{tab:1}, for the same degrees of freedom, TMM's memory consumption is approximately four times that of CRM.




\begin{table}[htbp!]
\renewcommand\arraystretch{1.5}
\centering
\caption{\label{tab:1}\inset{Comparison of the memory costs for the three methods, growing with $O[N^2]$, $O[(2N)^2]$ and $O[(lN)^2]$, respectively}}
\begin{tabular}{c|c|c|c|c}
\hline
\hline 
Unit cell unknowns ($N$)  &  500         &  1000       &  1500       &  2000      \\ \hline
Memory for CRM (Mb)       &  3.82       &  15.29       &  34.4       &  61.16     \\ \hline
Memory for TMM (Mb)       &  15.23      &  61.04       &  139.25     &  244.36    \\ \hline
Memory for SCM (Mb)       &  309.43     &  1238.22     &  2785.78    &  4953.98   \\ 
\hline
\hline
\end{tabular}
\footnotetext[1]{\inset{The SCM requires $l=9$ unit cells to achieve a convergence error of 1E-4 when $\eta=\omega/100$.}}
\end{table}

In terms of computation time, CRM is again more advantageous. Firstly, similar to the principle of memory consumption, the time consumption of the TMM primarily stems from the eigenanalysis of matrix $\bm{T}$, while that of the CRM mainly comes from the inversion of the coupling matrix $\bm{\zeta}^s$. As $\bm{T}$ has twice the degrees of freedom as $\bm{\zeta}^s$, the time consumption of the TMM is therefore greater. Secondly, although the time complexities for both the eigenanalysis and inversion of equivalent matrices are $O(N^3)$, the proportionality constant for eigenanalysis is larger, leading to a longer computation time for the TMM. Table~\ref{tab:2} presents the computation time required for both methods, showing that the TMM takes significantly longer than the CRM for the same number of degrees of freedom. \inset{It should be noted that the CRM time in Table~\ref{tab:2} also has to take into account the number of iterations.} Selecting an appropriate imaginary frequency $\eta$ can effectively reduce the iterations, but at the expense of decreased computational accuracy, which is discussed in~\cite{sha2021surface}.


\begin{table}[htbp!]
\renewcommand\arraystretch{1.5}
\centering
\caption{\label{tab:2}\inset{Comparison of the time costs for the three methods, growing with $O[N^3]$, $O[(2N)^3]$ and $O[(lN)^3]$, respectively}}
\begin{tabular}{c|c|c|c|c}
\hline
\hline
Unit cell unknowns ($N$)   &  500                    &  1000                 &  1500                   &  2000                \\ \hline
Time for CRM (s)           &  0.006 $\times$ 4       &  0.036 $\times$ 4      &  0.11 $\times$ 4       &  0.26 $\times$ 4   \\ \hline
Time for TMM (s)           &  0.63                   &  5.02                 &  18.27                  &  45.83              \\ \hline
Time for SCM (s)           &  2.34                   &  21.31                &  69.94                  &  167.71              \\ 
\hline
\hline
\end{tabular}
\footnotetext[1]{MATLAB run using an Intel Core i7-12700 CPU (12 cores, 20 threads).}
\footnotetext[2]{\inset{The CRM requires 4 iterations to achieve a convergence error of 1E-4 when $\eta=\omega/100$.}}
\end{table}


\inset{It is worth summarizing the respective advantages of CRM and TMM, together with the widely used SCM. CRM is most efficient in terms of time and memory assumption. Meanwhile, although SCM has slower convergence and thus requires more substantial computational resources, it can directly access eigenmode profiles [as shown in Fig.~\ref{fig2}(a)-Fig.~\ref{fig7}(a) calculated from the SCM]. Besides, despite that TMM consumes more time and memory when calculating surface Green's functions compared to CRM, it also offers certain advantages in other aspects. For example, the intermediate variables (eigenvalues $\bm{\Lambda}$ and eigenvectors $\bm{S}$) can be used to construct the surface bands (surface eigenvalues spectrum~\cite{mustonen2024exact}).}

\section{\label{sec:V}Discussion}
In this article, we apply two efficient algorithms to calculate the SDOS in photonic and acoustic crystals and investigate the corresponding topological phenomenon. The CRM focuses on efficient solving by reducing computational complexity using Gaussian elimination, while the TMM relies on direct solving through eigenanalysis of a transfer matrix. We provide numerical examples of various topological photonic and acoustic crystals to demonstrate the utility of these methods. 
\inset{The algorithms provided here enable the calculation of SDOS of inherently continuous systems that require discretized grid descriptions, mitigating the calculation of supercells with substantially larger numerical degrees of freedom. 
This feature makes the algorithms particularly effective for structures where boundary effects converge more slowly with increasing system size, such as Moir\'{e} superlattice photonic crystals and metasurfaces with slow-varying phase gradients.
Aside from the numerical efficiency described above, other unique advantages of these methods include the straightforward sorting of the boundary mode of a particular interface from bulk modes that may spectrally overlap and the direct comparison of SDOS spectra with near-field scanning experimental data.}

In addition to the photonic and acoustic systems addressed here, our approach can be similarly applied to other platforms such as circuits, as long as the eigenmatrices (or Hamiltonians) of the system exhibit a similar cyclic tridiagonal form [see Eq.~\eqref{eq:eigenmatrix0}] for periodic structures.
The algorithms can also be further extended and applied to non-Hermitian systems, where the calculation expression for LDOS defined by Green's functions [see Eq.~\eqref{eq:LDOS}] can remain valid if the left and right eigenmodes satisfy the completeness relation~\cite{chew2019green}. 
In this case, the algorithms presented in this paper can still be employed for fast evaluation of SDOS. For instance, the algorithm designed for sandwiched structure [as shown in Fig.~\ref{fig1}(d)] can be used to find the bound states in the continuum in a photonic crystal slab~\cite{hsu2016bound}.

Our method can also be generalized for higher-order topological systems~\cite{yue2019symmetry}. For example, regarding corner states in 2D systems, one can consider a structure that extends infinitely in the $\bm{a}$ and $\bm{b}$ directions but is finite in the $\bm{a}+\bm{b}$ direction, where $\bm{a}$ and $\bm{b}$ are primitive vectors. By taking a supercell in the $\bm{a}+\bm{b}$ direction as a new unit cell, we can apply the algorithm designed for heterostructures [as shown in Fig.~\ref{fig1}(c)] to obtain the corner-density-of-states spectrum.

\section{\label{apdx}Methods}
\inset{In the main text, we have showcased the method for calculating the surface Green's function using a bare semi-infinite structure. Here, we additionally supplement the methods for handling other complex cases and provide detailed explanations of a sandwiched structure as an example [see Fig.~\ref{fig1}(d)].}

\inset{In CRM [see Algorithm \ref{algorithm: CRM sandwich}], we define ${\bm{\alpha}_0}$, ${\bm{\beta}_0}$ and ${\bm{\zeta}_0}$ as the bulk couplings of a crystal on one side, and ${\bm{\overbar{\alpha}}_0}$, ${\bm{\overbar{\beta}}_0}$ and ${\bm{\overbar{\zeta}}_0}$ as the bulk couplings on the other side. We also take ${\bm{\alpha}^s_0}$, ${\bm{\beta}^s_0}$, ${\bm{\overbar{\alpha}}^s_0}$, ${\bm{\overbar{\beta}}^s_0}$ and ${\bm{\zeta}^s_0}$ as the interface couplings for the sandwiched crystal slab. Here, the overline $-$ and superscript $s$ denotes quantities in the opposite direction and at the interface, respectively, and the subscript $i$ indicates the iteration steps. Then we can independently iterate the bulk couplings in both directions, and update the interface couplings according to those of bulks. Finally, the surface Green's function can be found through $\bm{\zeta}^s_i$ if the convergence residual $(\bm{\zeta} _{i }^s-\bm{\zeta} _{i-1 }^s)/ (\bm{\zeta} _{i-1 }^s)$ is small enough. }

\inset{In TMM [see Algorithm \ref{algorithm: TMM sandwich}], we define $\bm{T}_1$, $\bm{T}_2$ and $\bm{\overbar{T}}_1$, $\bm{\overbar{T}}_2$ as the transfer matrices of crystals on two sides. Then we can do generalized eigenanalyses of the two pairs of transfer matrices independently. By ordering the corresponding eigenvalues and eigenvectors, and combining it with the equation of Green's function at the interface, we can also find exact solution of surface Green's function.}
\\

\begin{algorithm}[H]
    \SetAlgoLined
    {\bf Initialization:} \\
    maximum iterations $>$ 0, error tolerance $>$ 0\\
    ${\bm{\alpha} _0} = {\bm{Z}_{0,1}},\ {\bm{\beta} _0} = {\bm{Z}_{1,0}},\ {\bm{\zeta} _0} = {\bm{Z}_{0,0}}$\\
    $\bm{\zeta} _0^s = {\bm{Z}_{0,0}}$\\
        
    \For{$i=1$ {\rm \textbf{to}} {\rm maximum iterations}}{
    \inset{\texttt{ $// $ Parameters for the bulk crystal }} \\
    ${\bm{\alpha} _{i }} = {\bm{\alpha} _{i-1}}{\left( {{\bm{\zeta} _{i-1}}} \right)^{ - 1}}{\bm{\alpha} _{i-1}}$\\
    ${\bm{\beta} _{i }} =  {\bm{\beta} _{i-1}}{\left( {{\bm{\zeta} _{i-1}}} \right)^{ - 1}}{\bm{\beta} _{i-1}}$\\
    ${\bm{\zeta} _{i }} =  {\bm{\zeta} _{i-1}} - {\bm{\alpha} _{i-1}}{\left( {{\bm{\zeta} _{i-1}}} \right)^{ - 1}}{\bm{\beta} _{i-1}} - {\bm{\beta} _{i-1}}{\left( {{\bm{\zeta} _{i-1}}} \right)^{ - 1}}{\bm{\alpha} _{i-1}}$\\
    \inset{\texttt{ $// $ Parameters for the surface layer }} \\
    $\bm{\zeta} _{i }^s = \bm{\zeta} _{i-1}^s - \bm{\alpha} _{i-1}{\left( {{\bm{\zeta} _{i-1}}} \right)^{ - 1}}\bm{\beta} _{i-1}$\\
    \If {$(\bm{\zeta} _{i }^s-\bm{\zeta} _{i-1 }^s)/ (\bm{\zeta} _{i-1 }^s)$ $<$ {\rm error tolerance}}{\bf break}{}
    }
    ${\bm{G}_{0,0}} =  {\left( {\bm{\zeta} _{i }^s} \right)^{ - 1}}$
    \caption{CRM for a bare semi-infinite crystal}
    \label{algorithm: CRM bare}
\end{algorithm}

\begin{algorithm}[H]
    \SetAlgoLined
    {\bf Initialization:} \\
    maximum iterations $>$ 0, error tolerance $>$ 0\\
    ${\bm{\alpha} _0} = {\bm{Z}_{1,2}},\ {\bm{\beta} _0} = {\bm{Z}_{2,1}},\ {\bm{\zeta} _0} = {\bm{Z}_{1,1}}$\\
    $\bm{\alpha} _0^s = {\bm{Z}_{0,1}},\ \bm{\beta} _0^s = {\bm{Z}_{1,0}},\ \bm{\zeta} _0^s = {\bm{Z}_{0,0}} $  \\
    \For{$i=1$ {\rm \textbf{to}} {\rm maximum iterations}}{
    \inset{\texttt{ $// $ Parameters for the bulk crystal }} \\
    ${\bm{\alpha} _{i }} = {\bm{\alpha} _{i-1}}{\left( {{\bm{\zeta} _{i-1}}} \right)^{ - 1}}{\bm{\alpha} _{i-1}}$\\
    ${\bm{\beta} _{i }} =  {\bm{\beta} _{i-1}}{\left( {{\bm{\zeta} _{i-1}}} \right)^{ - 1}}{\bm{\beta} _{i-1}}$\\
    ${\bm{\zeta} _{i }} =  {\bm{\zeta} _{i-1}} - {\bm{\alpha} _{i-1}}{\left( {{\bm{\zeta} _{i-1}}} \right)^{ - 1}}{\bm{\beta} _{i-1}} - {\bm{\beta} _{i-1}}{\left( {{\bm{\zeta} _{i-1}}} \right)^{ - 1}}{\bm{\alpha} _{i-1}}$\\
    \inset{\texttt{ $// $ Parameters for the surface layer }} \\
    $\bm{\alpha} _{i }^s = \bm{\alpha} _{i-1}^s{\left( {{\bm{\zeta} _{i-1}}} \right)^{ - 1}}{\bm{\alpha} _{i-1}} $\\
    $\bm{\beta} _{i }^s =  \bm{\beta} _{i-1}{\left( {{\bm{\zeta} _{i-1}}} \right)^{ - 1}}{\bm{\beta} _{i-1}^s} $\\
    $\bm{\zeta} _{i }^s =  \bm{\zeta} _{i-1}^s - \bm{\alpha} _{i-1}^s{\left( {{\bm{\zeta} _{i-1}}} \right)^{ - 1}}\bm{\beta} _{i-1}^s$\\
    \If {$(\bm{\zeta} _{i }^s-\bm{\zeta} _{i-1 }^s)/ (\bm{\zeta} _{i-1 }^s)$ $<$ {\rm error tolerance}}{\bf break}{}
    }
    ${\bm{G}_{0,0}} =  {\left( {\bm{\zeta} _{i }^s} \right)^{ - 1}}$
    \caption{CRM for a semi-infinite coated crystal}
    \label{algorithm: CRM coated}
\end{algorithm}

\newpage 

\begin{minipage}[!t]{0.97\linewidth}
\begin{algorithm}[H]
    \SetAlgoLined
    {\bf Initialization:} \\
    maximum iterations $>$ 0, error tolerance $>$ 0\\
    ${\bm{\alpha} _0} = {\bm{Z}_{0,1}},\ {\bm{\beta} _0} = {\bm{Z}_{1,0}},\ {\bm{\zeta} _0} = {\bm{Z}_{0,0}}$\\
    ${\bm{\overbar{\alpha}} _0} = {\overbar{\bm{Z}}_{{1},{2}}},\ {\bm{\overbar{\beta}} _0} = {\overbar{\bm{Z}}_{{2},{1}}},\ \bm{\overbar{\zeta}} _0 = {\overbar{\bm{Z}}_{{1},{1}}}$\\
    $ \bm{\overbar{\alpha}} _0^s = {\overbar{\bm{Z}}_{{0},1}},\ \bm{\overbar{\beta}} _0^s = {\overbar{\bm{Z}}_{1,{0}}},\ \bm{\zeta} _0^s = {\bm{Z}_{0,0}}$\\
    \For{$i=1$ {\rm \textbf{to}} {\rm maximum iterations}}{
    \inset{\texttt{ $// $ Parameters for the bulk crystal 1 }} \\
${\bm{\alpha} _{i }} = {\bm{\alpha} _{i-1}}{\left( {{\bm{\zeta} _{i-1}}} \right)^{ - 1}}{\bm{\alpha} _{i-1}} $\\
${\bm{\beta} _{i }} =  {\bm{\beta} _{i-1}}{\left( {{\bm{\zeta} _{i-1}}} \right)^{ - 1}}{\bm{\beta} _{i-1}} $\\
${\bm{\zeta} _{i }} =  {\bm{\zeta} _{i-1}} - {\bm{\alpha} _{i-1}}{\left( {{\bm{\zeta} _{i-1}}} \right)^{ - 1}}{\bm{\beta} _{i-1}} - {\bm{\beta} _{i-1}}{\left( {{\bm{\zeta} _{i-1}}} \right)^{ - 1}}{\bm{\alpha} _{i-1}}$\\
    \inset{\texttt{ $// $ Parameters for the bulk crystal 2 }} \\
${\bm{\overbar{\alpha}} _{i }} =  {\bm{\overbar{\alpha}} _{i-1}}{\left( {{\bm{\overbar{\zeta}} _{i-1}}} \right)^{ - 1}}{\bm{\overbar{\alpha}} _{i-1}}$\\
${\bm{\overbar{\beta}} _{i }} =  {\bm{\overbar{\beta}} _{i-1}}{\left( {{\bm{\overbar{\zeta}} _{i-1}}} \right)^{ - 1}}{\bm{\overbar{\beta}} _{i-1}} $\\
${\bm{\overbar{\zeta}} _{i }} =  {\bm{\overbar{\zeta}} _{i-1}} - {\bm{\overbar{\alpha}} _{i-1}}{\left( {{\bm{\overbar{\zeta}} _{i-1}}} \right)^{ - 1}}{\bm{\overbar{\beta}} _{i-1}} - {\bm{\overbar{\beta}} _{i-1}}{\left( {{\bm{\overbar{\zeta}} _{i-1}}} \right)^{ - 1}}{\bm{\overbar{\alpha}} _{i-1}}$\\
    \inset{\texttt{ $// $ Parameters for the surface layer }}\\
$\bm{\overbar{\alpha}} _{i }^s =  \bm{\overbar{\alpha}} _{i-1}^s{\left( {{\bm{\overbar{\zeta}} _{i-1}}} \right)^{ - 1}}{\bm{\overbar{\alpha}} _{i-1}}$\\
$\bm{\overbar{\beta}} _{i }^s =  \bm{\overbar{\beta}} _{i-1}{\left( {{\bm{\overbar{\zeta}} _{i-1}}} \right)^{ - 1}}{\bm{\overbar{\beta}} _{i-1}^s} $\\
$\bm{\zeta} _{i }^s =  \bm{\zeta} _{i-1}^s - \bm{\alpha} _{i-1}{\left( {{\bm{\zeta} _{i-1}}} \right)^{ - 1}}\bm{\beta} _{i-1} -\bm{\overbar{\alpha}} _{i-1}^s{\left( {{\bm{\overbar{\zeta}} _{i-1}}} \right)^{ - 1}}\bm{\overbar{\beta}} _{i-1}^s$\\
    \If {$(\bm{\zeta} _{i }^s-\bm{\zeta} _{i-1 }^s)/ (\bm{\zeta} _{i-1 }^s)$ $<$ {\rm error tolerance}}{\bf break}{}
    }
    ${\bm{G}_{0,0}} =  {\left( {\bm{\zeta} _{i }^s} \right)^{ - 1}}$
    \caption{CRM for two semi-infinite crystals interfaced with each other}
    \label{algorithm: CRM hetero}
\end{algorithm}
\end{minipage}

\newpage 

\begin{algorithm}[H]
    \SetAlgoLined
    {\bf Initialization:} \\
    maximum iterations $>$ 0, error tolerance $>$ 0\\
    ${\bm{\alpha} _0} = {\bm{Z}_{1,2}},\ {\bm{\beta} _0} = {\bm{Z}_{2,1}},\ {\bm{\zeta} _0} = {\bm{Z}_{1,1}}$\\
    ${\bm{\overbar{\alpha}} _0} = {\overbar{\bm{Z}}_{{1},{2}}},\ {\bm{\overbar{\beta}} _0} = {\overbar{\bm{Z}}_{{2},{1}}},\ \bm{\overbar{\zeta}} _0 = {\overbar{\bm{Z}}_{{1},{1}}}$\\
    $ \bm{\alpha} _0^s = {\bm{Z}_{0,1}},\ \bm{\beta} _0^s = {\bm{Z}_{1,0}},\ $ \\
 $ \bm{\overbar{\alpha}} _0^s = {\overbar{\bm{Z}}_{{0},1}},\ \bm{\overbar{\beta}} _0^s = {\overbar{\bm{Z}}_{1,{0}}}, \bm{\zeta} _0^s = {\bm{Z}_{0,0}}$\\
    \For{$i=1$ {\rm \textbf{to}} {\rm maximum iterations}}{
    \inset{\texttt{ $// $ Parameters for the bulk crystal 1 }}\\
${\bm{\alpha} _{i }} = {\bm{\alpha} _{i-1}}{\left( {{\bm{\zeta} _{i-1}}} \right)^{ - 1}}{\bm{\alpha} _{i-1}} $\\
${\bm{\beta} _{i }} =  {\bm{\beta} _{i-1}}{\left( {{\bm{\zeta} _{i-1}}} \right)^{ - 1}}{\bm{\beta} _{i-1}} $\\
${\bm{\zeta} _{i }} =  {\bm{\zeta} _{i-1}} - {\bm{\alpha} _{i-1}}{\left( {{\bm{\zeta} _{i-1}}} \right)^{ - 1}}{\bm{\beta} _{i-1}} - {\bm{\beta} _{i-1}}{\left( {{\bm{\zeta} _{i-1}}} \right)^{ - 1}}{\bm{\alpha} _{i-1}}$\\
    \inset{\texttt{ $// $ Parameters for the bulk crystal 2 }}\\
${\bm{\overbar{\alpha}} _{i }} =  {\bm{\overbar{\alpha}} _{i-1}}{\left( {{\bm{\overbar{\zeta}} _{i-1}}} \right)^{ - 1}}{\bm{\overbar{\alpha}} _{i-1}} $\\
${\bm{\overbar{\beta}} _{i }} =  {\bm{\overbar{\beta}} _{i-1}}{\left( {{\bm{\overbar{\zeta}} _{i-1}}} \right)^{ - 1}}{\bm{\overbar{\beta}} _{i-1}} $\\
${\bm{\overbar{\zeta}} _{i }} =  {\bm{\overbar{\zeta}} _{i-1}} - {\bm{\overbar{\alpha}} _{i-1}}{\left( {{\bm{\overbar{\zeta}} _{i-1}}} \right)^{ - 1}}{\bm{\overbar{\beta}} _{i-1}} - {\bm{\overbar{\beta}} _{i-1}}{\left( {{\bm{\overbar{\zeta}} _{i-1}}} \right)^{ - 1}}{\bm{\overbar{\alpha}} _{i-1}}$\\
    \inset{\texttt{ $// $ Parameters for the surface layer }} \\
$\bm{\alpha} _{i }^s = \bm{\alpha} _{i-1}^s{\left( {{\bm{\zeta} _{i-1}}} \right)^{ - 1}}{\bm{\alpha} _{i-1}} $\\
$\bm{\beta} _{i }^s =  \bm{\beta} _{i-1}{\left( {{\bm{\zeta} _{i-1}}} \right)^{ - 1}}{\bm{\beta} _{i-1}^s} $\\
$\bm{\overbar{\alpha}} _{i }^s =  \bm{\overbar{\alpha}} _{i-1}^s{\left( {{\bm{\overbar{\zeta}} _{i-1}}} \right)^{ - 1}}{\bm{\overbar{\alpha}} _{i-1}} $\\
$\bm{\overbar{\beta}} _{i }^s =  \bm{\overbar{\beta}} _{i-1}{\left( {{\bm{\overbar{\zeta}} _{i-1}}} \right)^{ - 1}}{\bm{\overbar{\beta}} _{i-1}^s} $\\
$\bm{\zeta} _{i }^s =  \bm{\zeta} _{i-1}^s - \bm{\alpha} _{i-1}^s{\left( {{\bm{\zeta} _{i-1}}} \right)^{ - 1}}\bm{\beta} _{i-1}^s -\bm{\overbar{\alpha}} _{i-1}^s{\left( {{\bm{\overbar{\zeta}} _{i-1}}} \right)^{ - 1}}\bm{\overbar{\beta}} _{i-1}^s$\\
    \If {$(\bm{\zeta} _{i }^s-\bm{\zeta} _{i-1 }^s)/ (\bm{\zeta} _{i-1 }^s)$ $<$ {\rm error tolerance}}{\bf break}{}
    }
    ${\bm{G}_{0,0}} =  {\left( {\bm{\zeta} _{i }^s} \right)^{ - 1}}$
    \caption{CRM for two semi-infinite crystals separated by another crystal slab}
    \label{algorithm: CRM sandwich}
\end{algorithm}

\newpage 

\begin{algorithm}[H]
    \SetAlgoLined
    {\bf Initialization:} \\
    $\bm{T}_1=\begin{pmatrix}
    \bm{0}&\bm{I}\\
    -\bm{Z}_{0,1}&\bm{0}\\
    \end{pmatrix}$
    ,\ 
    $\bm{T}_2=\begin{pmatrix}
    \bm{I}&\bm{0}\\
    \bm{Z}_{0,0}&\bm{Z}_{1,0}\\
    \end{pmatrix}$\\
    Eigenanalysis $\bm{T_2}\bm{S}=\bm{T_1} \bm{S} \bm{\Lambda}$\\
    \inset{\texttt{ $/* $ Order the moduli of the eigenvalues }}\\
    \inset{\texttt{ $\bm{\Lambda}$ to obtain partitioned $\bm{S}=
    \begin{pmatrix}
    \bm{S}_1&\bm{S}_3\\
    \bm{S}_2&\bm{S}_4\\
    \end{pmatrix}$ 
    $*/ $ }}\\
    \For{$i=1$ {\rm \textbf{to}} {{\rm length} $(\bm{T}_1)$}}{
    \For{$j=$ {\rm length} $(\bm{T}_1)$ {\rm \textbf{to}} { $i+1$}}{
    \If {$\lvert \lambda(j) \rvert<\lvert \lambda(j-1) \rvert$}{
    Swap $\lambda(j)$ with $\lambda(j-1)$\\
    Swap $s(j)$ with $s(j-1)$\\
    }{}
    }
    }
    $\bm{G}_{0,0}=\left( {\bm{Z}_{0,0}} + {\bm{Z}_{0,1}} \bm{S_2} \bm{S_1}^{-1} \right) ^{-1}$
    \caption{TMM for a bare semi-infinite crystal}
    \label{algorithm: TMM bare}
\end{algorithm}

\begin{algorithm}[H]
    \SetAlgoLined
    {\bf Initialization:} \\
    $\bm{T}_1=\begin{pmatrix}
    \bm{0}&\bm{I}\\
    -\bm{Z}_{1,2}&\bm{0}\\
    \end{pmatrix}$
    ,\ 
    $\bm{T}_2=\begin{pmatrix}
    \bm{I}&\bm{0}\\
    \bm{Z}_{1,1}&\bm{Z}_{2,1}\\
    \end{pmatrix}$\\
    Eigenanalysis $\bm{T_2}\bm{S}=\bm{T_1} \bm{S} \bm{\Lambda}$\\
    \inset{\texttt{ $/* $ Order the moduli of the eigenvalues }}\\
    \inset{\texttt{ $\bm{\Lambda}$ to obtain partitioned $\bm{S}=
    \begin{pmatrix}
    \bm{S}_1&\bm{S}_3\\
    \bm{S}_2&\bm{S}_4\\
    \end{pmatrix}$ 
    $*/ $ }}\\
    \For{$i=1$ {\rm \textbf{to}} {{\rm length} $(\bm{T}_1)$}}{
    \For{$j=$ {\rm length} $(\bm{T}_1)$ {\rm \textbf{to}} { $i+1$}}{
    \If {$\lvert \lambda(j) \rvert<\lvert \lambda(j-1) \rvert$}{
    Swap $\lambda(j)$ with $\lambda(j-1)$\\
    Swap $s(j)$ with $s(j-1)$\\
    }{}
    }
    }
    $\bm{G}_{0,0}=\left[ {\bm{Z}_{0,0}} +{\bm{Z}_{0,1}} \left( {\bm{Z}_{1,1}}+ {\bm{Z}_{1,2}} \bm{S_2} \bm{S_1}^{-1} \right)^{-1}{\bm{Z}_{1,0}}\right] ^{-1}$
    \caption{TMM for a semi-infinite coated crystal}
    \label{algorithm: TMM coated}
\end{algorithm}

\newpage 

\begin{algorithm}[H]
    \SetAlgoLined
    {\bf Initialization:} \\
    $\bm{T}_1=\begin{pmatrix}
    \bm{0}&\bm{I}\\
    -\bm{Z}_{0,1}&\bm{0}\\
    \end{pmatrix}$
    ,\ 
    $\bm{T}_2=\begin{pmatrix}
    \bm{I}&\bm{0}\\
    \bm{Z}_{0,0}&\bm{Z}_{1,0}\\
    \end{pmatrix}$
    ,\
    \\
    $\overbar{\bm{T}}_1=\begin{pmatrix}
    \bm{0}&\bm{I}\\
    -\overbar{\bm{Z}}_{1,2}&\bm{0}
    \end{pmatrix}$
    ,\
    $\overbar{\bm{T}}_2=\begin{pmatrix}
    \bm{I}&\bm{0}\\
    \overbar{\bm{Z}}_{1,1}&\overbar{\bm{Z}}_{2,1}\\
    \end{pmatrix}$
    \\
    Eigenanalysis $\bm{T_2}\bm{S}=\bm{T_1} \bm{S} \bm{\Lambda}$\\
    \inset{\texttt{ $/* $ Order the moduli of the eigenvalues }}\\
    \inset{\texttt{ $\bm{\Lambda}$ to obtain partitioned $\bm{S}=
    \begin{pmatrix}
    \bm{S}_1&\bm{S}_3\\
    \bm{S}_2&\bm{S}_4\\
    \end{pmatrix}$ 
    $*/ $ }}\\
    \For{$i=1$ {\rm \textbf{to}} {{\rm length} $(\bm{T}_1)$}}{
    \For{$j=$ {\rm length} $(\bm{T}_1)$ {\rm \textbf{to}} { $i+1$}}{
    \If {$\lvert \lambda(j) \rvert<\lvert \lambda(j-1) \rvert$}{
    Swap $\lambda(j)$ with $\lambda(j-1)$\\
    Swap $s(j)$ with $s(j-1)$\\
    }{}
    }
    }
    Eigenanalysis $\overbar{\bm{T}}_2\bm{S}=\overbar{\bm{T}}_1 \overbar{\bm{S}} \overbar{\bm{\Lambda}}$\\
    \inset{ \texttt{ $/* $ Order the moduli of the eigenvalues }}\\
    \inset{ \texttt{ $\overbar{\bm{\Lambda}}$ to obtain partitioned $\overbar{\bm{S}}=
    \begin{pmatrix}
    \overbar{\bm{S}}_2 & \overbar{\bm{S}}_4\\
    \overbar{\bm{S}}_1 & \overbar{\bm{S}}_3\\
    \end{pmatrix}$ $*/$ }}\\
    \For{$i=1$ {\rm \textbf{to}} {{\rm length} $(\overbar{\bm{T}}_1)$}}{
    \For{$j=$ {\rm length} $(\overbar{\bm{T}}_1)$ {\rm \textbf{to}} { $i+1$}}{
    \If {$\lvert \overbar{\lambda}(j) \rvert<\lvert \overbar{\lambda}(j-1) \rvert$}{
    Swap $\overbar{\lambda}(j)$ with $\overbar{\lambda}(j-1)$\\
    Swap $\overbar{s}(j)$ with $\overbar{s}(j-1)$\\
    }{}
    }
    }
    $\bm{G}_{0,0}=\Big[ {\bm{Z}_{0,0}} +{\bm{Z}_{0,1}} \bm{S_2} \bm{S_1}^{-1}+$\\
    \qquad ${\overbar{\bm{Z}}_{0,{1}}} \left( {\overbar{\bm{Z}}_{{1},{1}}}+ 
    {\overbar{\bm{Z}}_{{1},{2}}} \bm{\overbar{S}_2} \bm{\overbar{S}_1}^{-1} \right)^{-1}{\overbar{\bm{Z}}_{{1},0}} \Big] ^{-1}$
    \caption{TMM for two semi-infinite crystals interfaced with each other}
    \label{algorithm: TMM hetero}
\end{algorithm}

\newpage 

\begin{algorithm}[H]
    \SetAlgoLined
    {\bf Initialization:} \\
    $\bm{T}_1=\begin{pmatrix}
    \bm{0}&\bm{I}\\
    -\bm{Z}_{1,2}&\bm{0}\\
    \end{pmatrix}$
    ,\ 
    $\bm{T}_2=\begin{pmatrix}
    \bm{I}&\bm{0}\\
    \bm{Z}_{1,1}&\bm{Z}_{2,1}\\
    \end{pmatrix}$
    ,\
    \\
    $\overbar{\bm{T}}_1=\begin{pmatrix}
    \bm{0}&\bm{I}\\
    -\overbar{\bm{Z}}_{1,2}&\bm{0}
    \end{pmatrix}$
    ,\
    $\overbar{\bm{T}}_2=\begin{pmatrix}
    \bm{I}&\bm{0}\\
    \overbar{\bm{Z}}_{1,1}&\overbar{\bm{Z}}_{2,1}\\
    \end{pmatrix}$
    \\
    Eigenanalysis $\bm{T_2}\bm{S}=\bm{T_1} \bm{S} \bm{\Lambda}$\\
    \inset{\texttt{ $/* $ Order the moduli of the eigenvalues }}\\
    \inset{\texttt{ $\bm{\Lambda}$ to obtain partitioned $\bm{S}=
    \begin{pmatrix}
    \bm{S}_1&\bm{S}_3\\
    \bm{S}_2&\bm{S}_4\\
    \end{pmatrix}$ 
    $*/ $ }}\\
    \For{$i=1$ {\rm \textbf{to}} {{\rm length} $(\bm{T}_1)$}}{
    \For{$j=$ {\rm length} $(\bm{T}_1)$ {\rm \textbf{to}} { $i+1$}}{
    \If {$\lvert \lambda(j) \rvert<\lvert \lambda(j-1) \rvert$}{
    Swap $\lambda(j)$ with $\lambda(j-1)$\\
    Swap $s(j)$ with $s(j-1)$\\
    }{}
    }
    }
    Eigenanalysis $\overbar{\bm{T}}_2\bm{S}=\overbar{\bm{T}}_1 \overbar{\bm{S}} \overbar{\bm{\Lambda}}$\\
    \inset{ \texttt{ $/* $ Order the moduli of the eigenvalues }}\\
    \inset{ \texttt{ $\overbar{\bm{\Lambda}}$ to obtain partitioned $\overbar{\bm{S}}=
    \begin{pmatrix}
    \overbar{\bm{S}}_2 & \overbar{\bm{S}}_4\\
    \overbar{\bm{S}}_1 & \overbar{\bm{S}}_3\\
    \end{pmatrix}$ $*/$ }}\\
    \For{$i=1$ {\rm \textbf{to}} {{\rm length} $(\overbar{\bm{T}}_1)$}}{
    \For{$j=$ {\rm length} $(\overbar{\bm{T}}_1)$ {\rm \textbf{to}} { $i+1$}}{
    \If {$\lvert \overbar{\lambda}(j) \rvert<\lvert \overbar{\lambda}(j-1) \rvert$}{
    Swap $\overbar{\lambda}(j)$ with $\overbar{\lambda}(j-1)$\\
    Swap $\overbar{s}(j)$ with $\overbar{s}(j-1)$\\
    }{}
    }
    }
    $\bm{G}_{0,0}=\Big[ {\bm{Z}_{0,0}} +{\bm{Z}_{0,1}} \left( {\bm{Z}_{1,1}}+ {\bm{Z}_{1,2}} \bm{S_2} \bm{S_1}^{-1} \right)^{-1}{\bm{Z}_{1,0}}
    +$\\
    \qquad $ {\overbar{\bm{Z}}_{0,{1}}} \left( {\overbar{\bm{Z}}_{{1},{1}}}+ {\overbar{\bm{Z}}_{{1},{2}}} \bm{\overbar{S}_2} \bm{\overbar{S}_1}^{-1} \right)^{-1}{\overbar{\bm{Z}}_{{1},0}} \Big] ^{-1}$
    \caption{TMM for two semi-infinite crystals separated by another crystal slab}
    \label{algorithm: TMM sandwich}
\end{algorithm}




\section{Acknowledgments}
The authors acknowledge the support from the National Natural Science Foundation of China Excellent Young Scientists Fund (12222417), the Hong Kong Research Grants Council through Early Career Scheme (27300924),  Strategic Topics (STG3/E-704/23-N), the Startup Fund of The University of Hong Kong, Ms.~Belinda Hung, the Asian Young Scientist Fellowship, the Croucher Foundation, and New Cornerstone Science Foundation through the Xplorer Prize. 
M. -Y. Xia acknowledges the support from the National Natural Science Foundation of China (62231001 and 62171005).
L. Lu acknowledges the support from the Natural Science Fund for Distinguish Young Scholars of China (12025409), the Beijing Natural Science Foundation (Z200008), and the Chinese Academy of Sciences through the Project for Young Scientists in Basic Research (YSBR-021).

\section{Author contributions}
Y.-X. Sha performed the research and drafted the manu-script. All authors discussed the results. Y.~Y. supervised the project.

\section{Competing interests}
The authors declare no competing interests.

%

\end{document}